

\def\up#1{\raise 1ex\hbox{\underbar{\sevenrm #1}}}
\def\cima#1{\raise 1ex\hbox{\sevenrm #1}}
\def\baixo#1{\kern-.125em \lower .6ex\hbox{\sevenrm #1}}
\def\cimaum#1{\hbox{\raise .5ex\hbox{$#1$}}}

\def\ii{\'{\i}}

\def\ao{\~ao\ }

\def\aov{\~ao,\ }

\def\ni{\noindent}

\def\ss{\smallskip}
\def\bs{\bigskip}
\def\ms{\medskip}

\font\eightrm=cmr8
\font\sevenrm=cmr7

 1
 1
 1


\magnification=\magstep1
\hsize=6truein
\baselineskip=16truept
\topskip 1cm
\magnification=\magstep1
\hbox to 6.5truein{\hfil DF/IST-3.92}
\hbox to 6.5truein{\hfil IFM-3/92}
\hbox to 6.5truein{\hfil DFFCUL 03-5/1992}

\vskip 1cm
\centerline{\bf  ON THE COSMOLOGY OF MASSIVE VECTOR FIELDS}
\vskip 0.5cm
\centerline{\bf WITH SO(3) GLOBAL SYMMETRY}
\vskip 1cm
\noindent
M.C. Bento$^{1)2)}$,
O. Bertolami$^{1)2)}$,
P.V. Moniz$^{2)3)}$\footnote{\dag}{Work supported in part by a JNICT
      graduate scholarship  BD/138/90-RM and by STRIDE/FEDER Project
      JN.91.02.},
J.M. Mour\ao$^{1)4)}$
and
P.M. S\'a$^{2)}$\footnote{\ddag}{Work supported by
        a GTAE (Grupo Te\'orico de Altas Energias) grant.}
\vskip 1cm
{\eightrm
\item{}
\itemitem{1)} Departamento de F\ii sica,
          Instituto Superior T\'ecnico,
          Av.\ Rovisco Pais 1,
          1096 Lisboa Codex, Portugal.
\vskip 0.2cm
\itemitem{2)} C.F.M.C.,
          Av.\ Prof.\ Gama Pinto 2,
          1699 Lisboa Codex, Portugal.
\vskip 0.2cm
\itemitem{3)} Departamento de F\ii sica,
          Faculdade Ci\^encias de Lisboa, Campo Grande C1-P4,
           1700 Lisboa Codex, Portugal.
\vskip 0.2cm
\itemitem{4)} C.F.N.,
          Av.\ Prof.\ Gama Pinto 2,
          1699 Lisboa Codex, Portugal.
}
\vskip 1.5cm
\centerline{\bf Abstract}
\vskip 0.5cm
We study the dynamics of flat Friedmann-Robertson-Walker (FRW)
cosmologies in
the presence of a triplet of  massive vector fields  with $SO(3)$ global
symmetry. We find an $E^3$-symmetric ansatz for the vector fields
that is compatible with the $E^3$-invariant FRW
metric and propose a method to make invariant ans\" atze for more
general cosmological models. We use techniques of dynamical systems to
study
qualitatively the behaviour of the model and find, in particular, that
the effective equation of state of the system changes gradually from a
radiation-dominated to a matter-dominated form and that the scale of the
transition depends on the mass of the gauge fields.

\vfill
\eject
\noindent
{\bf I. INTRODUCTION}

\vskip 0.5cm

Cosmological solutions to the Einstein-Yang-Mills (EYM) equations have
been known for some time. In particular,  solutions corresponding
to open, closed and  flat Friedmann-Robertson-Walker
(FRW) universes, in the case where the
gauge group is $SU(2)$, were studied in refs.\ [1,2].
The generalization of these results to
arbitrary gauge groups, for  closed FRW models, was done in
refs.\ [3,4] (the euclidian case),
 where an $SO(4)$-symmetric ansatz is derived from the theory of symmetric
fields on homogeneous spaces [5]. This $SO(4)$-symmetric
ansatz has been also used  for constructing the
wave-function of a Universe dominated by radiation [6]
and to study cosmological implications of
generalized Kaluza-Klein theories
[7]. The case of a flat FRW
inflationary model in the presence of $E^3$-symmetric gauge and scalar
fields was studied in ref.\ [8].

In treating the classical dynamics of the very early universe
with
the help of a variational principle applied to actions derived
from particle physics theories, one hopes to obtain a better
understanding of the effective equations of state valid in the classical
regime. On the other hand, this approach may be useful in order to
bridge the gap between the rather
distinct descriptions of the classical and quantum regimes that have
been  used sofar. In
this context, it is quite natural from the particle physics viewpoint
to study non-abelian gauge theories in a cosmological setting.
In realistic particle-physics-motivated cosmological models, however,
the vector bosons normally acquire a mass thereby breaking
the gauge symmetry. As this process takes place in the very early
Universe, it is of
interest to investigate cosmological solutions of the resulting theory.

The purpose of this paper is  to investigate homogeneous and
isotropic solutions to the EYM system after symmetry breaking. We shall
 assume that
the gauge symmetry one starts with is $SO_I(3)$, which subsequently
breaks down  to a global $SO_I(3)$
 (through a mechanism which we choose not to specify
but can be thought of as the usual Higgs mechanism) and that, in this
process, the gauge fields acquire a mass.
We derive an
$E^3$-symmetric ansatz for the triplet of massive vector fields with
$SO_I(3)$
global symmetry that is compatible with the $E^3$-invariant (flat) FRW
ansatz for the metric. We compare the dynamical behaviour of the system
before and after symmetry breaking and find, in
particular, that, in the
latter phase, the equation of state changes  with time, gradually from a
radiation-dominated to a matter-dominated form
and that the scale of the transition depends on the mass of the vector
field.

Homogeneous and isotropic solutions exist for other gauge groups and
symmetry breaking patterns (see Appendix).
For the case of electroweak symmetry breaking,
$SU(2)\otimes U(1)$  down to a $U(1)$ however,
at least for the ansatz proposed below,
homogeneity and isotropy cannot be achieved,
as the vector fields acquire different masses after symmetry
breaking thus preventing the possibility for the
energy-momentum tensor (EMT) to have the perfect fluid form.

 Recently, there has been
great interest in models where inflation is driven by massive vector
fields [9]. In these works, the vector field is a $U(1)$ gauge
field which, upon symmetry breaking, acquires a potential leading to
inflation. Such a vector field is not allowed by an isotropic
spacetime and consequently the authors have to choose an anisotropic
cosmological model (Bianchi I), which, at least
in some cases, leads to excessive anisotropy at
late times. The present work shows that if one starts from a non-abelian
gauge theory this problem may not arise since, in this case, it is in
general possible to built models which are compatible with an FRW
cosmology from the start. However, the presence of a mass term does not by
itself lead to
sufficient inflation and more involved potentials have to be used in
order to achieve it [9,10]. In this paper, we shall be interested only
in
analysing the consequences of a mass term for the vector fields and
therefore we do not find inflationary solutions.

The paper is organized as follows. In section 2, we derive the
$E^3$-invariant ansatz for a triplet of massive vector fields with
$SO_I(3)$ global symmetry. In section
3, we briefly discuss
non-trivial solutions to the coupled EYM equations with gauge group
$SO_I(3)$ which correspond to the spatially flat FRW universe. We then
examine the case where the vector fields become massive
and derive the corresponding equations of motion from an effective
action built upon
substitution of the $E^3$-symmetric ansatz in the initial action. In
section 4, we analyse the solutions of the
equations of motion using  methods of the theory of
dynamical systems and in section 5 we present our conclusions.
Finally, in the Appendix, we present a  method to find
$G$-symmetric ans\" atze (where $G$ is the isometry group) for massive
vector fields with an arbitrary global internal symmetry.
\bs
\noindent
{\bf \item{II.} ANSATZ FOR VECTOR FIELDS WITH SO(3) GLOBAL INTERNAL
SYMMETRY}
\vskip 0.5cm
The spacetime in flat FRW universes has the form

$$M^4=R^4=R\times E^3/SO(3),\eqno(1)$$

\noindent
where the six-dimensional euclidean group $E^3$ is the isometry group of
the spatial hypersurfaces

$$M^3=R^3=E^3/SO(3).\eqno(2)$$
\ni
The isotropy group $SO(3)$ leaves the origin $o\in R^3=E^3/SO(3)$
invariant but rotates the
curves which pass through $o$ and therefore defines a representation of
$SO(3)$ in the  tangent space at the origin $T_o R^3$. This is the
so-called isotropy representation which, in this case,
is just the vector representation $\underline 3$ of $SO(3)$.

The fact that there are no singlets in the isotropy
representation has far reaching implications.
Indeed, if a  vector field $X$ in $R^3$ is  $E^3$-invariant then,
in particular, it must be
invariant at $o$, i.e., a singlet under the action of $SO(3)$ on $T_o R^3$.
Since this is not
possible, we conclude that there are no $E^3$-invariant vector fields in
$R^3$.

Being physically obvious for $R^3$ (as it just means that there are no
homogeneous and isotropic vector fields) this result is
important as it applies for any homogeneous space $G/H$ for which
the isotropy
representation has no trivial singlets. This is also the case for the
spatial hypersurfaces $S^3$ and $H^3$ of closed and open FRW models
which, therefore, do not have vector fields invariant under the
action of the corresponding isometry group either. This  explains
why
a single vector field is incompatible with FRW geometry; indeed, being
non-spatially homogeneous and isotropic, the vector fields generate
anisotropic energy-momentum tensors which therefore cannot be
proportional to the Einstein tensor of the FRW metric.
\par
In this section, we shall study the case of flat FRW models, leaving the
analysis of the more general
case of a (not necessarily four-dimensional) spacetime of the form
$M^d=R\times G/H$ to the Appendix. Models with $d>4$ are
relevant to
the study of the cosmological implications of Kaluza-Klein theories [7].
\par
If, instead of a single vector field, we consider a theory with a
multiplet
of (covariant) vector fields $A_\mu^a, a=1,...,N$, where $a$ is an
internal
index, we may hope that, although the EMT corresponding to each field
$A_\mu^a$ is not $E^3$-invariant, it would be possible to relate the
asymmetries of the different vector fields in such a way that the
total EMT is $E^3$-invariant. Let us then consider the case of a triplet
of massive vector fields with global $SO_I(3)$ symmetry.
The generators of the isometry group $G=E^3$ satisfy the
following commutation relations

$$\eqalign{ \left[ T_i,T_j\right] &=0, \cr
            \left[ Q_i,Q_j\right] &=\epsilon_{ijk} Q_k,\cr
            \left[ Q_i,T_j\right] &=\epsilon_{ijk}T_k;\qquad i,j,k=1,2,3
\cr}\eqno(3)
 $$

\noindent
so that the $T_i$ span an (abelian) invariant subalgebra of
Lie($E^3$) --- the
algebra of translations. The corresponding Killing vector fields are

$$X_i={\partial \over \partial x^i},\eqno(4)  $$

\noindent
where
$x^i$ are the cartesian coordinates in $R^3$  and
$\partial\over {\partial x^i}$
together with ${\partial \over \partial t}$ form a (global)
 moving frame in $R^4=R\times E^3/SO(3), \{X_\mu
\}=\{{\partial\over \partial t},{\partial \over \partial
x^i}\}$, with the dual coframe being given by
$\{\omega^\mu\}=\{dt,dx^i\}$ ($\mu=0,1,2,3$).
The remaining Killing vector fields, corresponding to the rotations
$Q_i$, read

$$Y_i=-\epsilon_{ijk}x^j {\partial \over \partial x^k},
\eqno(5)
$$

\noindent
so that

$$\eqalignno{ {\cal L}_{X_i}X_j &=[X_i,X_j]=0, & (6-a)\cr
              {\cal L}_{Y_i}X_j &=-{\cal
L}_{X_j}Y_i=[Y_i,X_j]=\epsilon_{ijk}X_k, & (6-b) \cr
              {\cal L}_{Y_i}Y_j &=[Y_i,Y_j]=\epsilon_{ijk}Y_k. & (6-c)
\cr
} $$

The conditions of spatial homogeneity and isotropy read, infinitesimaly

$$\eqalignno{{\cal L}_{X_i}A& =0,&(7-a) \cr
           {\cal L}_{Y_i}A&=0,& (7-b)\cr}
 $$

\noindent
where $A=A_\mu^a(t,\vec x) \omega^\mu L_a$, the $L_a$ being the
generators of the internal group $SO_I(3)$

$$\left[L_a,L_b\right]=\epsilon_{abc}L_c. \eqno(8)  $$
\ni
In agreement with our previous  considerations, we conclude that the
only solution of
(7)
is the trivial one, i.e. such that only the zeroth component of the
covector fields is different from zero: $A=A_o^a(t) dt L_a$.
Condition (7-b) is, however, too restrictive and we now consider a
weaker version of it by allowing the field $A$ to be
such that infinitesimal
spatial rotations  generate internal $SO_I(3)$ rotations

$$\eqalignno{{\cal L}_{X_i}A& =0, &(9-a)\cr
           {\cal L}_{Y_i}A&=-\left[ L_i,A\right].&(9-b)\cr}$$
\ni
It is clear that, if (9) takes place, then, although $A$ is not
$E^3$-invariant, it generates an EMT that is  invariant under the action
of
$E^3$. Indeed, the action of $E^3$ in the EMT is equivalent to an
internal rotation with respect to which the  EMT is invariant

$${\cal L}_{Y_i}EMT(A)=-\delta_{L_i}EMT(A)=0.
\eqno(10)$$
\ni
 The conditions for spatial homogeneity, eq.\ (9-a), imply that the
components
$A_\mu^a$  depend only on  the time variable $t$. In order to
solve (9-b), let us consider the linear space of vector fields with
constant coefficients: $C=\{X= a^\mu X_\mu, a^\mu \in R\}$.
Eq.\ (6-b) implies that $X \in C$ transforms as $\underline 1 \oplus
\underline
3$
under $SO(3)$ spatial rotations. On the other hand, the one-form $A$ can be
considered, for any $t$, as a linear mapping from $C$ to the internal space
(i.e. to the Lie algebra of $SO_I(3)$)

$$A(X)=A(a^\mu  X_\mu)=A_\mu^a(t) a^\mu L_a,\eqno(11)$$

\noindent
in which the triplet representation $\underline 3$ of $SO_I(3)$ is
realized. Rewritting (9-b) in the form

$$A({\cal L}_{Y_i}X)=[L_i,A(X)],  \eqno(12)$$

\noindent
we see that $A$ intertwines the representations
$\underline 1 \oplus \underline
3$ and
$\underline 3$ of $SO(3)$ and $SO_I(3) {isom\atop \cong}SO(3)$,
respectively.
Schur's lemma then implies that $A$ vanishes between subspaces of
non-equivalent representations

$$A\left( {\partial \over \partial
t}\right)=0\Longleftrightarrow A_o=0,\eqno(13)$$

\noindent
and is proportional to the identity operator between subspaces with
equivalent representations

$$A\left({\partial \over \partial
x_i}\right)=\chi_o(t)L_i\Longleftrightarrow
           A_i^a(t) =\chi_o(t)\delta_i^a,\eqno(14) $$

\ni
where $\chi_o(t)$ is an arbitrary function.
 Alternatively, the $E^3$-symmetric ansatz for $A$ satisfying (9) and
therefore generating $E^3$-invariant EMT, can be written as

$$A=\chi_o(t) dx^i L_i.\eqno(15)$$.

\bs
\def\os{s\kern-3.5truemm \odot}
\noindent
{\bf III. EFFECTIVE ACTION AND FIELD EQUATIONS}
\vskip 0.2cm
\noindent
{\bf A.  Massless case }
\vskip 0.5cm
We start with the following action for the coupled EYM system with
$SO_I(3)$ gauge group

$$S=\int_{M^4} d^4 x \sqrt{-g} \left({ 1 \over 2k^2}R+{1\over 8 e^2}
Tr \left[F_{\mu\nu}F^{\mu\nu}\right]\right),
\eqno(16)$$

\noindent
where $k^2=8\pi G$ and $e$ is the gauge coupling. Our conventions
 and notation for the gravitational
part of the action correspond to those of refs.\ [3,4]. The gauge field
strenght is $F_{\mu
\nu}=\partial_{\mu}A_{\nu}-\partial_{\nu}A_{\mu}+[A_{\mu},A_{\nu}]$,
where  $A_{\mu}=A_{\mu}^{a}L_{a}$.
We assume that the spacetime manifold $M^4$ is topologically of the type~(1),
where $\{t_o\} \times E^3/SO(3)=\{t_0\} \times R^3, t_o\in R$,
are the spatial orbits of the isometry group
$G=E^3=R^3 \os SO(3)$.

The FRW ansatz, for flat models, gives the most general form of an
$E^3$-invariant metric

$$ds^2=-N^2(t) dt^2+a^2(t) \left( (dx^1)^2+(dx^2)^2+(dx^3)^2
                           \right) \eqno(17)$$

\noindent
where the lapse
function $N(t)$ and the scale factor $a(t)$ are arbitrary non-vanishing
functions.

Cosmological models
in the presence of gauge fields have been investigated in some detail in
the literature
[1-3,7,8]; for these models, the
local symmetry of the action allows  the  use of
the theory of symmetric gauge fields [5] in order to make the necessary
ans\"atze. The $G$-symmetric ansatz
corresponds
to fields which, under the action of the isometry group $G$, undergo
a local internal rotation; however,  in theories
with global
symmetry such fields do not, in general, lead to $G$-invariant EMTs, as
can be seen from eq.\ (10) (for more details, the reader is referred
to the Appendix).
  Ansatz (15), for (covariant) vector fields with global
$SO_I(3)$ internal symmetry is, of course, also valid in the case of
action (16), with local $SO_I(3)$ symmetry, and we shall use it
in what follows.
It is easy to check that this ansatz leads to a (traceless)
energy-momentum tensor (EMT) with the required $E^3$-invariance

$${T_o}^o=- \hat \rho(t),\qquad {T_i}^j=\hat p(t) {\delta_i}^j,
\eqno(18)
$$

\noindent
i.e., it corresponds to the EMT of a perfect comoving
fluid with energy density $\hat \rho$ and pressure $\hat p$ given by

$$\eqalign{\hat \rho & = {3\over 4a^2 e^2 N}\left[ {\dot
                                            \chi_{o}^{2}\over
                    2}+{N^2\over a^2}V_{gf}\right], \cr
           \hat  p   & = {1\over 4a^2 e^2 N}\left[ {\dot
                                            \chi_{o}^{2}\over
                    2}+{N^2\over a^2}V_{gf}\right], \cr
              V_{gf} & = {1\over 8}\chi_{o}^{4},  \cr
}\eqno(19)  $$

\noindent
where the dot denotes the derivative with respect to $t$; moreover,
the equation of state for the system is clearly that of a radiation
fluid:
$\hat \rho=3 \hat p$.

There are, for this model, two equivalent methods to obtain
 the dynamical equations for
the functions  $a(t)$ and $\chi_{o}$(t): one can either substitute
the ans\"atze, eqs.\ (13), (14) and (17), into the field equations
derived from action
(16) or substitute the ans\"atze directly into the action and then get
the
dynamical equations from this effective action. The latter method has
the advantage of simplifying the study of the equations of motion and we
shall adopt it in what follows. The effective action is given by
(disregarding the infinite volume of the spatial hypersurface)

$$S_{eff}=\int^{t_{2}}_{t_{1}} dt \left[ -{3\over k^2}
{{\dot a}^2 a\over N}+
{3a\over
4 e^2N}\left(
 {\dot \chi_{o}^{2}\over 2}-{N^2\over a^2}V_{gf}  \right) \right].
\eqno(20)$$

\ni
The equations of motion are obtained upon variation of this action with
respect to $N(t)$, $a(t)$, and $\chi_{o}$(t). We get, respectively, in the
``gauge'' $N=1$

$$\eqalignno{           H^2  &={k^2 \over 3}\hat\rho, & (21-a)\cr
                \dot H +H^2  &= -{k^2\over 6}(\hat\rho+3\hat p),
&(21-b)\cr
\ddot \chi_{o}+\dot \chi_{o} H &= -{1\over 2a^2} \chi_{o}^{3},
&(21-c)\cr
}   $$

\noindent
with $\hat \rho$ and $\hat p$ given by eq.\ (19) and $H\equiv\dot a /a$.

The Yang-Mills equation, (21-c), takes a much simpler form in the
conformal time [3,4]

$$d\eta = {dt\over a(t)},\eqno(22)$$

\noindent
since, in this case, this equation decouples from the remaining ones. Indeed,
substituting (22) in (21-c), we obtain

$${\chi_o}^{\prime \prime}= - {\chi_o^3\over 2},\eqno(23)$$

\noindent
where the prime denotes the derivative with respect to the conformal
time.
 The solution of
eq.\ (23) is given by

$$\chi_{o}=f(\eta;E,\eta_{0}),\eqno(24)$$

\noindent
where $f=f(\eta;E,\eta_{0})$ is a function defined implicitly by

$$\eta-\eta_{0}= {\left( 2E \right)}^{-1/4} F(\gamma(f,E),r).
\eqno(25)$$
\ni
In the above equation, $F(x,k)$ denotes the elliptic function of the
first kind [11],

$$\eqalign{\gamma(f,E) &=\cos^{-1}\left[ {\left( 8E
                         \right)}^{-1/4} f\right], \cr
                  r &={\sqrt{2}\over 2}, \cr }\eqno(26)
 $$

\noindent
and the constant $E$ is the conserved ``mechanical energy'' of the
solution

$$ E={1\over 2} (\chi^{\prime}_{o})^{2}+V_{gf}. \eqno(27)$$

\noindent
Using a ``mechanical'' analogy, we can consider eq.\ (23) as describing
the motion of a particle with kinetic energy ${\chi_o^{\prime 2} \over
2}
$, potential energy $V_{gf}$ and total energy $E$. The above solution
 describes therefore a particle oscillating in a potential well
between the turning
points:   $-(8E)^{1/4}$ and $(8E)^{1/4}$.
\vskip 0.2cm
\ni
{\bf B. Massive case}
\vskip 0.5cm
 Consider now the action which results from adding a mass
term for the gauge fields [12] to the action of eq.\ (16)

$$S'=
\int_{M^4} d^4 x \sqrt{-g} \left({ 1 \over 2k^2}R+{1\over 8 e^2}
Tr \left[F_{\mu\nu}F^{\mu\nu}\right]+
 { 1\over 2}\ m^2 Tr[A_\mu A^\mu]\right). \eqno(28)
$$
\ni
Such a mass term would arise, e.g., via the Higgs mechanism, with a
complex
doublet of scalar particles acquiring a vacuum expectation value.
The above action is no longer gauge invariant
 and only a global $SO_I(3)$ symmetry remains. Substituting
 the ans\"atze, eqs.\ (13), (14) and (17), into the
energy-momentum tensor corresponding to the action (28), we obtain

$$\eqalign{ {T_o}^o&=- \hat\rho-{3\over 4}m^2 \left({\chi_o\over
a}\right)^2\equiv -\rho, \cr
            {T_i}^j&=\left(\hat p-{1\over 4}m^2
\left({\chi_o\over
a}\right)^2 \right)
{\delta_i}^j\equiv  p\ {\delta_i^j},
\cr
}\eqno(29)$$

\noindent
with $\hat\rho$ and $\hat p$ given by eq.\ (19), which shows that the
perfect fluid form of the EMT is preserved in spite
of the presence of the mass term. Notice that, if instead of a mass term
 we had a quartic potential, this could be absorbed in a
redefinition of $V_{gf}$, which would lead us back to the results of
 the previous subsection.

We are interested in spatially homogeneous and isotropic cosmologies
associated with the action (28).
Hence, we restrict ourselves to the set of $E^3$-symmetric
configurations for the fields $g_{\mu \nu}$ and $A_{\mu}$.
Substituting these ans\"atze into (28), we obtain for the effective action

$$S_{eff}=\int^{t_{2}}_{t_{1}} dt \left[- {3\over k^2}
{{\dot a}^2 a\over N}+
{3a\over
4 e^2N}\left(
 {\dot \chi_{o}^{2}\over 2}-{N^2\over a^2}V_{gf}  \right)
-{3\over 4} N m^2 {\chi_o}^2 a\right].
\eqno(30)$$

\ni
The equations of motion, derived upon variation with respect to $N(t)$,
$a(t)$ and $\chi_o$(t) are, respectively, in the ``gauge'' $N=1$:

$$\eqalignno{           H^2  &={k^2 \over 3} \rho ,&(31-a)\cr
                \dot H +H^2  &= -{k^2\over 6}(\rho + 3 p),
&(31-b)\cr
\ddot \chi_{o}+\dot \chi_{o} H &= -{1\over 2a^2} \chi_{o}^{3}-2m^2e^2\chi_o,
&(31-c)\cr
}   $$

\noindent
with $\rho$ and $p$ as given by eq.\ (29). Consider now the following
change of variables

$$\eqalign{x& =k {\chi_o\over a},\cr
           y&={k^2\over 2 \sqrt{2}}{\dot \chi_o\over a},\cr
           z& =k H,\cr
           \tau&=t/k.\cr }\eqno(32)$$
\ni
The system of eqs.\ (31), in the new (dimensionless) variables, can be
written as

$$\eqalignno{x_\tau &=  2\sqrt{2} y
                       -xz,                          &(33-a)\cr
             y_\tau &= -{\sqrt{2} \over 2} \mu^2 x
                       -2yz
                       -{\sqrt{2} \over 8} x^3,      &(33-b)\cr
             z_\tau &= -2 z^2
                       +{1 \over 4} \mu^2 x^2,
                                                     &(33-c)\cr
                z^2 &=  y^2
                       +{1\over 32} x^4
                       +{1\over 4} \mu^2 x^2,        &(33-d)\cr} $$

\noindent
where $\mu^2=m^2 k^2$ and we have set $e=1$; the index $\tau$ denotes
the derivative
with respect to $\tau$. Eqs.\ (33-a)--(33-c) define a three-dimensional
dynamical system
in the variables $x$, $y$, $z$ and eq.\ (33-d) is a constraint equation which
defines, in
$R^3$, the phase space
of the dynamical system.
Substituting the constraint into  eqs. (33-a), (33-b), we
are left with a two-dimensional dynamical system in the variables
$x,y$

$$\eqalign{x_\tau &= 2\sqrt{2} y
                   - x  \left(  y^2
                              + {1\over 32} x^4
                              + {1\over 4} \mu^2 x^2
                        \right)^{1\over 2},\cr
           y_\tau &=-{{\sqrt{2}}\over 2}\mu^2 x
                    -2y \left( y^2
                                              +{1\over {32}} x^4
                                              +{1\over 4} \mu^2 x^2
                                        \right)^{1\over 2}
                    - {{\sqrt{2}}\over 8}x^3,\cr
}\eqno(34)$$

\noindent
where we have considered the positive root of the constraint (33-d)
 as we are interested  in expanding models.

 It is this system of differential equations (34) that we are going to study
using the methods of the qualitative theory of dynamical systems [13].
\vskip 0.5cm
\noindent
{\bf IV.  QUALITATIVE ANALYSIS}
\vskip 0.5cm

The dynamical system (34) has just one critical point in the finite
region of variation of $x,y$, the origin, hereon referred to as $F(0,0)$
or simply $F$.
The remaining critical points, if any, will lie at infinity. In order to
study them, we complete the phase space with an infinitely distant
boundary (for which $x^2 + y^2 =+\infty$). It is also convenient to
perform a change of variables to polar coordinates, $(x,y) \rightarrow
(r,\theta)$, followed by a compactification of the entire phase space
into a circle of unit radius. This can be achieved through the
introduction of yet new radial and time coordinates $\rho$ and $\zeta$,
such that

$$r={\rho \over 1-\rho}\qquad (0\leq \rho \leq 1);\qquad  {d\zeta \over
d\tau}={1 \over (1-\rho)^2}.\eqno(35)
$$
\ni
In the variables $(\rho,\theta,\zeta)$, the system (34) becomes

$$\eqalign{ \rho_\zeta & =\sqrt{2}(2-{\mu^2 \over
2})\rho(1-\rho)^3\sin\theta\cos\theta-\rho^2(1-\rho)
(1-\sin^2\theta)f(\rho,\theta)\cr
             \phantom{\rho_\zeta =}& -{\sqrt{2}\over 8}\rho^3
(1-\rho) \sin \theta \cos^3 \theta \equiv
\Pi(\rho,\theta),\cr
            \theta_\zeta & = -\sqrt{2}
(1-\rho)^2\left(2\sin^2\theta+{\mu^2\over
2}
\cos^2\theta\right)-{\sqrt{2}
\over 8}\rho^2\cos^4\theta-\rho\sin\theta\cos\theta f(\rho,\theta)\cr
               \phantom{\theta_\zeta} &\equiv
\Psi(\rho,\theta),\cr}\eqno(36)
$$

\noindent
where $f^2(\rho,\theta)=(1-\rho)^2\sin^2\theta+{\rho^2\over
32}\cos^4\theta+{\mu^2\over 4}(1-\rho)^2\cos^2\theta$.

The critical points in the infinitely distant boundary $(\rho=1)$ are
the solutions of the equations: $\Pi(1,\theta)=0$ and
$\Psi(1,\theta)=0$, namely,
$N_1(1,{\pi\over 2})$,
$N_2(1,{{3\pi}\over 2})$,
$S_1(1,{{3\pi}\over 4})$,
$S_2(1,{{7\pi}\over 4})$ ---
see Fig.1.

We now turn to the analysis of the nature of the critical points of the
system. First, we would like to point out that, since (36) is invariant
under $\theta \rightarrow \theta+\pi$, we need only consider two of the
four critical points at infinity; hence, we are
left  with three points to analyse, $F(0,0)$ and, e.g.,
$N_1(1,{\pi\over 2})$
 and $S_1(1,{{3\pi}\over 4})$.

Regarding $F(0,0)$, it is easy to check that this point is degenerate
(see Table 1) and therefore the Hartman-Grobman theorem [13] does not
apply. We resort to the method known
as
``blow-up''
[13], which allows us to
establish that $F$ is a focus. In order to investigate its stability,
 we use Liapounov's theorem [13]. Consider the Liapounov
function

$${\cal L}(x,y)=2 (y^2 + {1\over {32}} x^4 + {{\mu^2}\over 4}
x^2).\eqno(37)
$$
\ni
Since ${\cal L}(x,y)$ obeys the conditions

$$\eqalign{ {\cal L}(0,0) &=0, \qquad {\cal L}((x,y)\not=(0,0)) >0, \cr
             {d{\cal L} \over d\zeta} & < 0 \qquad  \hbox{\rm in}\qquad
R^2/ \{ (0,0) \},
\cr}\eqno(38)
$$

\noindent
we conclude that $F$ is a stable focus and all
trajectories in the $x,y$ plane asymptotically approach this point.
Furthermore, trajectories in the neighbourhood of $F$ are
clockwise directed spirals.

Regarding the critical point at infinity $S_1$, the linear approximation of the
system (36) around this point is sufficient
and we conclude that $S_1$ is a saddle point (cf. Table 1).
As for $N_1$, the situation is similar after we perform a change of time
variable
($\zeta \rightarrow \hat \zeta$) such that ${d \hat \zeta \over
d\zeta}=(\rho-1)$ and we conclude that
$ N_1$ is an improper
unstable node (cf. Table 1). We shall also use the fact that the
trajectories in the vicinity of
$N_1$ are tangent
to the eigenvector which corresponds to the lowest eigenvalue ---
see Table 1 and Fig. 1.

In order to get a more detailed picture of phase space, we divide it
into
four distinct regions and find approximate analytic expressions for the
trajectories in each of these regions, namely

$$ \eqalign{\hbox{a)}\quad x&={2\over \mu }{1\over {\tau +C_1}}\cos
\tau\cr
                 y&={1\over \tau+C_1}\sin\tau     \cr
             \hbox{b)}\quad x&+y=2C_2 x^2    \cr
             \hbox{c)}\quad y&=x+C_3      \cr
              \hbox{d)}\quad y&=C_4 x^2\cr}
\eqalign{
\phantom{\mid x\mid} &\phantom{\simeq\mid y\mid;\mid x\mid,\mid y\mid \gg
1} \cr
\phantom{\mid x\mid} &\phantom{\simeq\mid y\mid;\mid x\mid,\mid y\mid \gg
1} \cr
          \mid x\mid & \ll 1;\mid y\mid\ll 1\cr
    \mid x\mid &\simeq \mid y\mid \gg 1
                                                              \cr
    \mid x\mid &\gg\mid y\mid\gg 1\cr
          \mid y\mid &\gg x^2;\mid  x\mid,\mid y\mid \gg
1\cr}\eqno(39)$$

\vskip 0.2cm
\noindent
where the $C_i\ (i=1,...,4)$ are arbitrary constants.
In case a), i.e., near the origin, we see that the shape of the
trajectories depends on the value of $\mu^2$. In fact, for small
values of $\mu^2$ the trajectories wind around F with an oblate shape
whereas for large values of $\mu^2$ they have a prolate shape. It is
indeed in this region that the behaviour
of the system is most sensitive to the presence of the mass term.

We now turn our attention to the analysis of the effective equation
of state for the system near the critical points. In the vicinity of
$S_1$, $\theta\simeq {3\pi\over 4}$; introducing this
approximation into the dynamical system, we get the following solutions

$$a=a_o t^{1/2},\qquad \chi_o=-{2 a_o\over
{(2k^2)}^{1/4}}\qquad (t\rightarrow 0)\eqno(40)$$
\ni
Eq.\ (29) then implies $\rho\simeq 3 p$, indicating that the effective
equation of state near $S_1$ is approximately that of a radiation fluid.
In fact, this is true everywhere except in the vicinity of $F$. This is to
be expected since, were it not for the mass term, the equation of state
would be $\rho=3 p$  throughout all regions of phase space and, as we
have already
mentioned, it is near the origin that the mass term most affects the
trajectories.

The effective equation of state near the focus can be analysed using a
procedure similar to the one explained above for
$S_1$, which leads us to conclude that

$$ a\sim t^{2/3},\qquad
   \chi_o \sim t^{-1/3} \sin {t\over \sqrt{k}} \qquad
   (t\rightarrow +\infty),\eqno(41)$$

\noindent
implying that $p\simeq 0$, corresponding to the equation of state for
dust matter.

In addition, we find that the time of transition between
the regimes described by the
different equations of state depends on the value of $m^2$ such that
the transiton occurs earlier for more massive vector fields. Before the
transition the energy density and  pressure evolve as expected in a
radiation-dominated Universe and, as the system approaches the focus $F$,
the Universe becomes matter-dominated and the pressure starts to
oscillate around zero (see Fig. 2)
with ever decreasing amplitude and quickly vanishes.
\vskip 1cm
\noindent
{\bf 5. CONCLUSIONS}
\vskip 0.5cm
We have shown that the Einstein-Yang-Mills theory, with gauge group
$SO_I(3)$, broken by the presence of a mass term for the gauge fields in
such a way that only a global $SO_I(3)$ symmetry remains, admits
homogeneous and isotropic flat FRW solutions. To prove this, we
have constructed an $E^3$-symmetric ansatz for the (massive) vector
fields,
 parametrized in terms of an arbitrary function of time $\chi_o(t)$.

 We have studied the equations of motion of this theory
using the methods of the theory of dynamical systems and found that the
system
has just one stable critical point, the origin, which is a focus. We
have checked that the behaviour of
the system is basically the same as for the unbroken phase except near
the origin;
in this region, the change is such that the effective
equation of state of the system
becomes that of a matter-dominated universe\footnote{$^1$}
{When the present paper
was already in press we became aware of ref.
[14] where an exact cosmological
solution for an isotropic gas of collisionless massive particles has been
found.
This solution also describes a gradual change of the equation of state
from $p = 1/3 \rho$ to $p = 0$.}.
\bs

\noindent
{\bf APPENDIX}
\vskip 0.5cm
Let us consider a spacetime of the form

$$M^d=R\times G/H,\eqno(A.1)$$

\noindent
where $\{t_o \}\times G/H$ are spatial hypersurfaces, $G$ is the isometry
group and $t\in R$ is a time-like coordinate. There are two classes of
models for
which vector fields can be compatible with the above geometry, in the
sense that solutions of the coupled Einstein-vector field equations
can be found in a rather simple way: non-abelian gauge theories and
theories with a global internal symmetry.
In the former case, since the
theory is invariant under an internal  local symmetry,  the
total EMT is also gauge  invariant, which implies that, for
the total EMT to be $G$-invariant, the fields $A_\mu^a $ need only to be
$G$-invariant up to a gauge transformation.  For a gauge group $K$
containing an $H$ subgroup such fields often exist, which has allowed the
study
of FRW cosmologies in the presence of non-abelian gauge sectors [3,7,8].
The
latter case has, however, to our knowledge, not yet been studied in the
literature. In section 2, we analysed a particular example, with
$G/H=E^3/SO(3)$ and $K=SO_I(3)$. We now
study the general case.

 Consider configurations
$A_\mu^a
(x)
$ for which the action of
$g\in G$ can be compensated by an internal global $K$-rotation

$$g^*\left(A_\mu^a (x) dx^\mu\right)=\Psi(\lambda(g^{-1}))_b^a
A_\mu^b(x) dx^\mu,\eqno(A.2))$$

\noindent
where $\lambda:G\rightarrow K $ is an homomorphism and $\Psi$ is the
representation of
$K$ in the internal space. Clearly, configurations satisfying (A.2) will
generate $G$-invariant (spatially homogeneous and isotropic in the FRW
case) EMTs. Notice that condition (A.2) is less restrictive than the
one of
$G$-invariance (corresponding in (A.2) to $\lambda(g)=e,\forall g \in  G$)
but still
much more restrictive than the one for gauge theories since the
compensating internal rotation, $\Psi (\lambda(g))$, cannot depend on
the points $x \in R\times G/H$.
Nevertheless, we will show  that, for a class of homogeneous spaces $G/H$
which includes the spatial hypersurfaces of closed and flat FRW models,
field configurations satisfying (A.2) can be found.

Let $\{ T_{\hbox{\^ \i }}\}_{\hbox{\^ \i } =1}^{dim G} $ be a basis in
Lie($G$), the Lie algebra of the isometry group $G$, satisfying:
$\left[T_{\hbox{\^ \i }},T_{\hbox{\^ \j}}\right]=C_{\hbox{\^ \i  \^
\j}}^{\hbox{\^ k}} T_{\hbox{\^
k}}$ and $\{X_{\hbox{\^ \i }}\} $ be the associated Killing vector
fields

$$X_{\hbox{\^ \i }}(x)={d\over d\epsilon}\left( exp(-\epsilon\
T_{\hbox{\^ \i }})x\right)\vert_{\epsilon=0}.\eqno(A.3)
$$
\ni
We shall  assume that the homogeneous space $G/H$ is such that
 there exists in Lie($G$) an
invariant subalgebra
(ideal) with basis $\{ T_{i}\}_{ i =1}^{dim G/H} $

$$\left[T_{\hbox{\^ \i }},T_{j}\right]=C_{\hbox{\^ \i j}}^{ k}
T_{k},\eqno(A.4)$$

\noindent
and such that $\{X_{i}\}_{i=1}^{dim G/H}$
 is a (local) moving frame in $G/H$. This is obviously the case
for flat (see sect.\ 2) and closed FRW models. Consider now the moving
frame in
$R\times G/H,\
\{X_{\mu}\}_{\mu=0}^{dim G/H}=\{{\partial \over
\partial t},X_{i}\}_{i=1}^{dim G/H}$,
and the moving coframe
$\{\omega^{\mu}\}_{\mu=0}^{dim G/H}=\{dt,\omega^i;i=1,...,dim G/H\}$
dual to $\{X_{\mu}\}_{\mu=0}^{dim G/H} $. The commutation
relations  define a representation of Lie($G$) acting in the
subalgebra of vector fields

$$C=\{X=a^\mu X_\mu, a^\mu \in R \},\eqno(A.5) $$

\noindent
according to

$${\cal L}_{X_{\hbox{\^ \i }}}X=a^{\prime \mu}X_\mu,\eqno(A.6)  $$

\noindent
where

$$a^{\prime o}=0, \quad a^{\prime i}=C^i_{k\hbox{\^ \i }}a^k. \eqno(A.7)
$$
\ni
In the coframe
$\{\omega^\mu\}$,
we  choose the fields $A_\mu^a$ to depend
 only on the coordinate $t$

$$A^a=A_\mu^a(t) \omega^\mu. \eqno(A.8) $$
\ni
As the $\omega^\mu $ are not $G$-invariant, the fields $A^a$  are  not
$G$-invariant either.
Infinitesimally,  condition (A.2) applied to (A.8) gives

$${\cal L}_{X_{\hbox{\^ \i }}}A^a=-\Psi\left( \lambda(T_{\hbox{\^
\i}})\right)_b^a A^b,\eqno(A.9)
$$
\noindent
or equivalently

$$A^a\left({\cal L}_{X_{\hbox{\^ \i }}}  X\right)=
\Psi\left( \lambda(T_{\hbox{\^ \i } })\right)_b^a A^b(X).\eqno(A.10)
 $$
\ni
These are linear algebraic constraints on $A_\mu^a(t) $, which imply
that, as a mapping from $C$ to the internal space (for fixed $t$)

$$A_\mu^a: \quad a^\mu \rightarrow B^a=A_\mu^a a^\mu, \eqno(A.11)  $$

\noindent
$A_\mu^a $ intertwines the representation of Lie($G$) in $C$ with the
representation of\break  $\lambda$(Lie($G$)) in the internal space.
{}From Schur's lemma,
we then conclude that there exist nontrivial solutions of (A.10) if
and only if there are equivalent irreducible representations in $C$ and in
the internal space. Therefore, in order
to make a $G$-symmetric ansatz for the fields $A_\mu^a$, one must:

\item{(i)} Check whether (A.4) takes place for the given $G$ and $G/H$
and choose the associated coframe $\{\omega^\mu\}$.

\item{(ii)} Choose an homomorphism $\lambda$ of the isometry group $G$ to
the internal symmetry group $K$ such that there are equivalent irreducible
representations in the representations of $G$ in $C$ (see (A.6), (A.7)) and
of
$\lambda(G)$ in the internal space.

\item{(iii)} Find the intertwining operators $A_\mu^a$ between the above
two representations and substitute them in (A.8).

\bs
{\bf REFERENCES}
\ms
\item{[1]} M. Henneaux, J. Math. Phys. {\bf 23}, 830 (1982).
\ss
\item{[2]} D.V. Galt'sov and M.S. Volkov, Phys. Lett. {\bf 256B}, 17
(1991).
\ss
\item{[3]} P.V. Moniz and J.M. Mour\aov Class. Quantum Grav. {\bf
8}, 1815 (1991).
\ss
\item{[4]} O. Bertolami, J.M. Mour\aov R.J. Picken and I.P. Volobuev,
Int. J. Mod. Phys. {\bf A6}, 4149 (1991).
\ss
\item{[5]} N.S. Manton, Nucl. Phys. {\bf B158}, 141 (1979);\par
           P. Forgacs and N.S. Manton, Commun. Math. Phys. {\bf 72}, 15
(1980);\par
\item{} R. Coquereuaux and A. Jadcyzk in ``{\it Riemannian Geometry,
Fiber Bundles,
Kaluza-Klein Theories and All That}'' -- Lectures Notes in Physics,
Vol.16
(World Scientific, Singapore, 1988);\par
\item{} Yu.A. Kubyshin, J.M. Mour\aov G. Rudolph and I.P. Volobuev in
``Dimensional Reduction of Gauge Theories, Spontaneous Compactification
and Model Building'' -- Lecture Notes in Physics, Vol. 349
(Springer, Berlin, 1989);
\item{} D. Kapetanakis, G. Zoupanos, ``{\it Coset Space Dimensional
Reduction of Gauge Theories'', Preprint CERN-TH 6359/91, MPI-Ph 92-29,
TUM-TH-140/92
}, to be published in {\it Phys. Rep.}
\ss
\item{[6]} O. Bertolami and J.M. Mour\aov Class. Quantum Grav.
{\bf
8}, 1271 (1991).
\ss
\item{[7]} Yu.A. Kubyshin, V.A. Rubakov and I.I. Tkachev, Int. J.
Mod. Phys. {\bf A4}, 1409 (1989); O. Bertolami, Yu.A. Kubyshin and J.M.
Mour\aov Phys. Rev.
{\bf D45}, 3405 (1992).
\ss
\item{[8]} P.V. Moniz, J.M. Mour\ao and P.M. S\'a, ``{\it The dynamics of
a flat Friedmann-Robertson-Walker inflationary model in the presence of
gauge fields}'', Lisbon Preprint IFM-3/91.
\ss
\item{[9]} L. H. Ford, Phys. Rev. {\bf D40}, 967 (1989);\par
           A.B. Burd and J.E. Lidsey, Nucl. Phys. {\bf B351}, 679
(1991).
\ss
\item{[10]} R.D. Ball and A.M. Matheson, Phys. Rev. {\bf
D45}, 2647 (1992).
\ss
\item{[11]} I.S. Gradshteyn and I.M. Ryzhik, {\it ``Tables of
Integrals, Series and Products''} (Academic Press, New York 1965).
\ss
\item{[12]} Vector fields in the presence of gravity have been
discussed, for example, by G. Tauber, J. Math. Phys. {\bf 10}, 33
(1969).
\ss
\item{[13]} M.W. Hirsch and S. Smale, {\it ``Differential Equations,
Dynamical
Systems and Linear Algebra''} (Academic Press, New York 1974);\par
\item{ } O.I. Bogoyavlensky, {\it ``Methods in the Qualitative Theory
of Dynamical
Systems in Astrophysics and Gas Dynamics''} (Springer, Berlin, 1985);\par
\item{} J. Carr, {\it ``Applications of Center Manifold Theory''}
(Springer, Berlin, 1981).
\item{[14]} H.H. Soleng, Astron. Atrophys. {\bf 237} (1990) 1.

\eject
\topskip 5cm
\moveright 0.5in
\vbox{\offinterlineskip
\halign{{\vrule height 16pt }\quad #\quad
&\vrule\hfil\quad
#\quad\hfil
&\vrule \hfil\quad #\quad\hfil &\vrule\hfil\quad #\quad\hfil\vrule\cr
\noalign{\hrule}
Point & Eigenvalues & Eigenvectors & Classification \cr
\noalign{\hrule}
$F(0,0)$ &$\pm i\sqrt{2} \mu$& & Focus
\cr
         &  &   &          \cr
 & & & \cr
\noalign{\hrule}
$S_1(1,{3\pi\over 4})$&${\sqrt{2}\over 16}$
&${\partial\over\partial\rho}$& Saddle\cr
                        &$-{\sqrt{2}\over
16}$&${\partial\over\partial\theta}$
&\cr
& & & \cr
\noalign{\hrule}
$N_1(1,{\pi\over 2})$&2 &${\partial
\over\partial\rho}-{\displaystyle 2\sqrt{2}}
{\partial\over\partial\theta}$& Unstable node\cr
           & 1 &${\partial\over \partial\theta}$ &   \cr
& & & \cr
\noalign{\hrule}
}}

\vskip 0.5cm
\item{{\bf Table 1.}} Classification of critical points from the
analysis of the dynamical system of eqs. (34) and (36).

\vfill
\eject

FIGURE CAPTIONS:
\vskip 1.5cm
\noindent
\item{Fig. 1 } All trajectories of the phase diagram of the model
asymptotically approach the focus $F$; $N_1,\ N_2,\,S_1,\ S_2$ are
critical points at infinity. Near the focus, trajectories change from
oblate to prolate shape as the mass of the vector bosons increases.

\vskip 0.5cm
\noindent
\item{Fig. 2} $3p/\rho$ as a function of time
showing the change of the equation of state of the Universe from
radiation-dominated to matter-dominated.

\bye